\documentclass[a4paper]{jpconf}
\usepackage{graphicx}
\usepackage{gensymb}
\begin{document}
\title{Moffatt vortices in the lid-driven cavity flow}

\author{Sougata Biswas$^\dagger$, Jiten C. Kalita$^\ddag$}

\address{Department of Mathematics, Indian Institute of Technology Guwahati, PIN 781039, India}

\ead{$^\dagger$b.sougata@iitg.ernet.in, $^\ddag$jiten@iitg.ernet.in}
\begin{abstract}
In incompressible viscous flows in a confined domain, vortices are known to form at the corners and in the vicinity of separation points. The existence of a sequence of vortices (known as Moffatt vortices) at the corner with diminishing size and rapidly decreasing intensity has been indicated by physical experiments as well as mathematical asymptotics. In this work, we establish the existence of Moffatt vortices for the flow in the famous Lid-driven square cavity at moderate Reynolds numbers by using an efficient Navier-Stokes solver on non-uniform space grids. We establish that Moffatt vortices in succession follow fixed geometric ratios in size and intensities for a particular Reynolds number. In order to eliminate the possibility of spurious solutions, we confirm the physical presence of the small scales by pressure gradient computation along the walls.
\end{abstract}
\section{Introduction}
In Stokes flow near a corner between two intersecting solid boundaries or a solid boundary and a free surface, the existence of a sequence of counter-rotating vortices was first established theoretically by H. K. Moffatt \cite{moff1,moff2}.
A flow visualization experiment by Taneda \cite{taneda} in a V-notch  endorsed the existence of these vortices. They are known as ``Moffatt vortices'', named aptly after H. K. Moffatt. Such vortices are formed due to rotation of a cylinder or external stirring force close to the solid boundaries. A careful undermining into the existing literature reveals that the study of Moffatt vortices in Stokes flow on different geometries \cite{anderson, kirkinis, chetan, malyuga, moff1, moff2, s1} has so far been established mostly through theoretical studies. There have been very few numerical studies \cite{col, s2} on this topic available in the existing literature. 
In this work, we study the existence of Moffatt eddies at the bottom corners for the flow in Lid-driven cavity at moderate Reynolds numbers by using a recently developed Higher-Order-Compact (HOC) scheme by Kalita \emph{et al.}  \cite{jiten} on non-uniform space grids. We further establish that the  Moffatt eddies in succession follow geometric ratios of the size and intensities which again depend on Reynolds numbers. By grid independence study and computing pressure gradients along the walls of the cavity, we confirm that the smallest scales captured by us are not numerical artifacts.

This paper is organized in four sections. The problem is briefly described in Section \ref{sec2}, while Section \ref{sec3} deals with numerical procedures and other related issues, Section \ref{sec4} with results and discussions and finally, in Section \ref{sec5} we summarize the whole work.
\section{Problem Description}\label{sec2}
\begin{figure}[hH]
\begin{center}
\includegraphics[height=6.6cm]{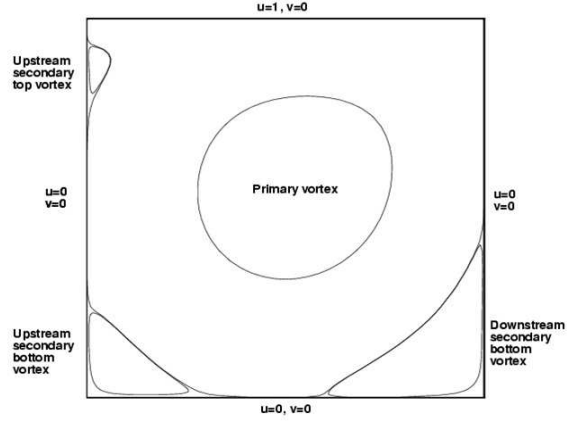}
\caption{Lid-driven square cavity.}\label{fig1}
\end{center}
\end{figure}

We have considered the classical Lid-driven square cavity problem (see Figure \ref{fig1}) which is one of the most extensively solved problems in CFD as it displays almost all fluid mechanical phenomena for incompressible viscous flows. 
The governing equations for the flow in cavity are the two-dimensional (2D) incompressible Navier-Stokes (N-S) equations which in the non-dimensional
primitive-variables form are given by
\begin{eqnarray}
 \frac{\partial u}{\partial x}+\frac{\partial v}{\partial y}=0,\hspace{3.8cm} \label{eqn1}\\
 \frac{\partial u}{\partial t}+u\frac{\partial u}{\partial x}+v\frac{\partial u}{\partial y}=-\frac{\partial p}{\partial x}+\frac{1}{Re}\nabla^2u,\label{eqn2} \\
 \frac{\partial v}{\partial t}+u\frac{\partial v}{\partial x}+v\frac{\partial v}{\partial y}=-\frac{\partial p}{\partial y}+\frac{1}{Re}\nabla^2v,\hspace{0.1cm}\label{eqn3}
\end{eqnarray}
Here $u$ and $v$ are the velocities along the $x$- and $y$- directions, $t$ is the time, $p$ is the pressure and $Re=\frac{UL}{\nu}$ is the Reynolds number where $U$ is the characteristic velocity (velocity of the lid), $L$ is the characteristic length (side of the square cavity) and $\nu$ is the kinematic viscosity.

Introducing streamfunction $\psi$ and vorticity $\omega$, the above equations (\ref{eqn1})-(\ref{eqn3}) transform to
\begin{eqnarray}
 \omega_t-\frac{1}{Re}(\omega_{xx}+\omega_{yy})+(u\omega_x+v\omega_y) =0,\label{eqn4}\\
 \psi_{xx}+\psi_{yy} = -\omega(x,y)\label{eqn5},\\
 \mbox{with} \quad u=\psi_y,\;v=-\psi_x \quad \mathrm{and} \quad \omega=v_x-u_y. \label{eqn6}
\end{eqnarray}
\section{Numerical Procedure and Other Related Issues}\label{sec3}
The unsteady convection-diffusion equation in 2D for a flow variable $\phi$ can be written as
\begin{equation}
 b\frac{\partial \phi}{\partial t}-\nabla^2\phi+c(x,y,t)\frac{\partial \phi}{\partial x}+d(x,y,t)\frac{\partial \phi}{\partial y}=f(x,y,t),\label{eqn7}
\end{equation}
where $b>0$ is a constant, $c$ and $d$ are the convection coefficients in the $x$- and $y$-directions, respectively, and $f$ is a forcing function.
The HOC scheme by Kalita \emph{et al.} \cite{jiten} for this equation on non-uniform space grid is given by:
\begin{eqnarray}\label{eqn8}
&&b[1+(H_1+H_2c)\delta_x+(K_1+K_2d)\delta_y+\{H_2-0.5(x_f-x_b)(H_1+H_2c)\}\delta_x^2\nonumber \\
&&+\{K_2-0.5(y_f-y_b)(K_1+K_2d)\}\delta_y^2]\delta_t^{+}\phi_{ij}^n\nonumber \\
&&+[-A_{ij}\delta_x^2-B_{ij}\delta_y^2+C_{ij}\delta_x+D_{ij}\delta_y+G_{ij}\delta_x\delta_y-H_{ij}\delta_x\delta_y^2
-K_{ij}\delta_x^2\delta_y-L_{ij}\delta_x^2\delta_y^2]\phi_{ij}^n\nonumber\\
&&=F_{ij},
\end{eqnarray}
where $\delta_t^{+}$ denotes the forward difference operator for time with uniform time step $\Delta t$ and $n$ represents the time level; the
coefficients $A_{ij}$, $B_{ij}$, $C_{ij}$, $D_{ij}$, $G_{ij}$, $H_{ij}$, $K_{ij}$, $L_{ij}$ and $F_{ij}$ can be found in Kalita \emph{et al.} \cite{jiten}.

In order to capture Moffatt vortices on relatively coarser grids, it is essential that the regions in the neighborhood of the solid boundaries are filled with cluster grids.
To generate a centro-symmetric grid with clustering near the walls, we use the stretching function \cite{jiten2}
\begin{equation}\label{eqn9}
 x_i=\frac{i}{i_{max}}-\frac{\lambda}{2\pi}\sin\bigg(\frac{2\pi i}{i_{max}}\bigg),\; 0<\lambda\leq 1
\end{equation}
in both $x$- and $y$-directions.

The boundary conditions for velocity on the top wall are given by $u=U=1$, $v=0$. On other walls of the cavity, the velocities are zero i.e, $u=0$, $v=0$. 
For streamfunction vorticity formulation, streamfunction values are zero along the four walls i.e, $\psi =0$. For
vorticity a transient HOC wall boundary approximation \cite{jiten} has been used.

In order to solve the algebraic systems associated with the finite difference approximation (\ref{eqn8}) the biconjugate gradient stabilized method (BiCGStab) \cite{jiten, kelly}
has been used because non-uniform grid invariably leads to non-symmetric matrices.
\section{Results and Discussions}\label{sec4}
For the problem under consideration, computations were carried out for $100 \leq Re \leq 3200$ on grids of sizes $81 \times 81$, $161 \times 161$ and $321 \times 321$.
\subsection{Evidence of Moffatt vortices in the cavity} 
\begin{figure}[hH]
(a)
\begin{minipage}{.5\linewidth}
\centering{\includegraphics[width=6cm]{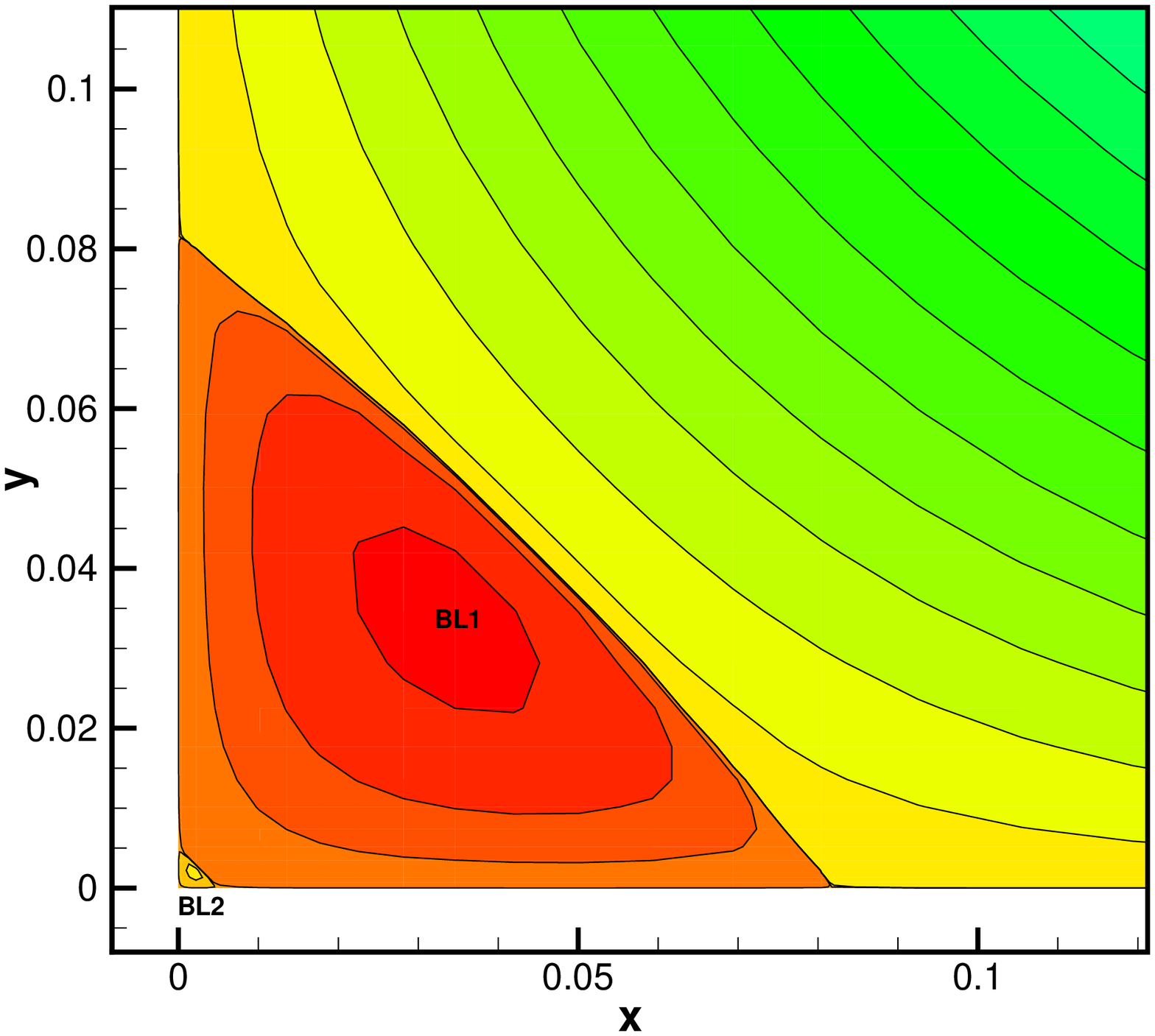}}
\end{minipage}
(b)
\begin{minipage}{.5\linewidth}
\centering{\includegraphics[width=6cm]{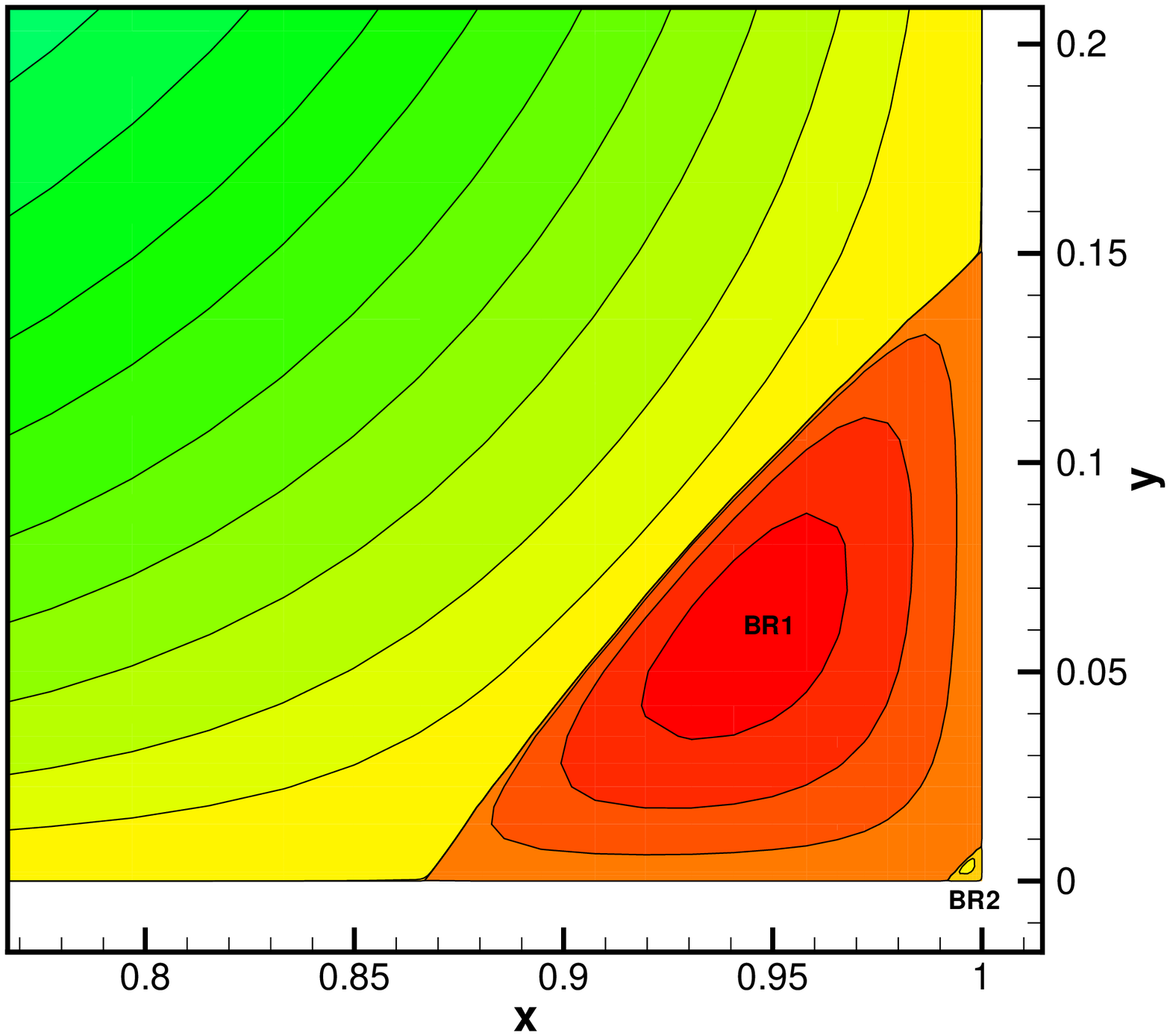}}
\end{minipage}
\caption{Moffatt vortices for $Re=100$ on grid size $81 \times 81$: (a) BL1, BL2 at bottom left corner and (b) BR1, BR2 at bottom right corner.}\label{fig2}
\end{figure}

\begin{figure}[hH]
\begin{minipage}{.5\linewidth}
\centering{\includegraphics[width=5.65cm]{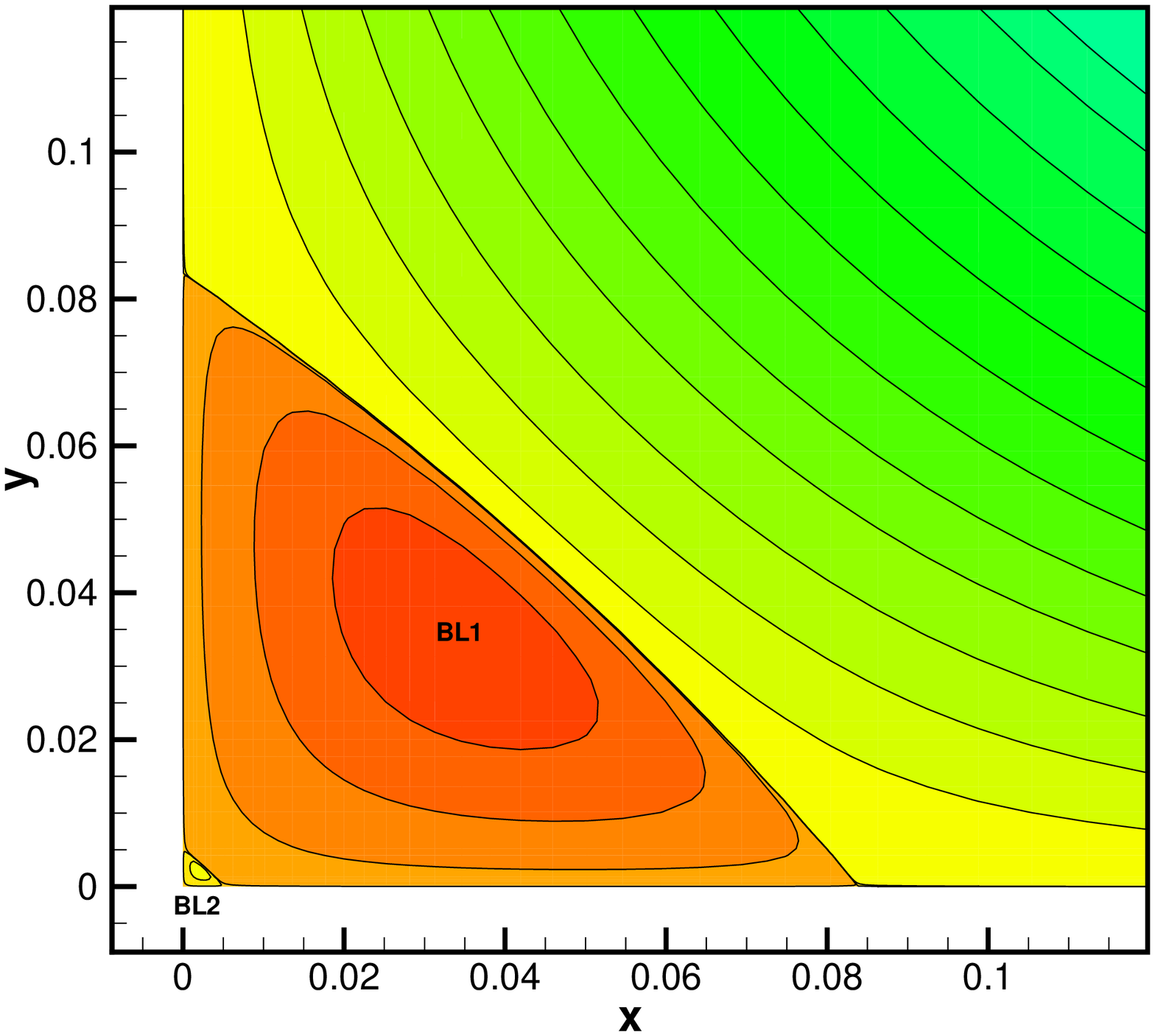}}
\end{minipage}
(a)
\begin{minipage}{.5\linewidth}
\centering{\includegraphics[width=5.65cm]{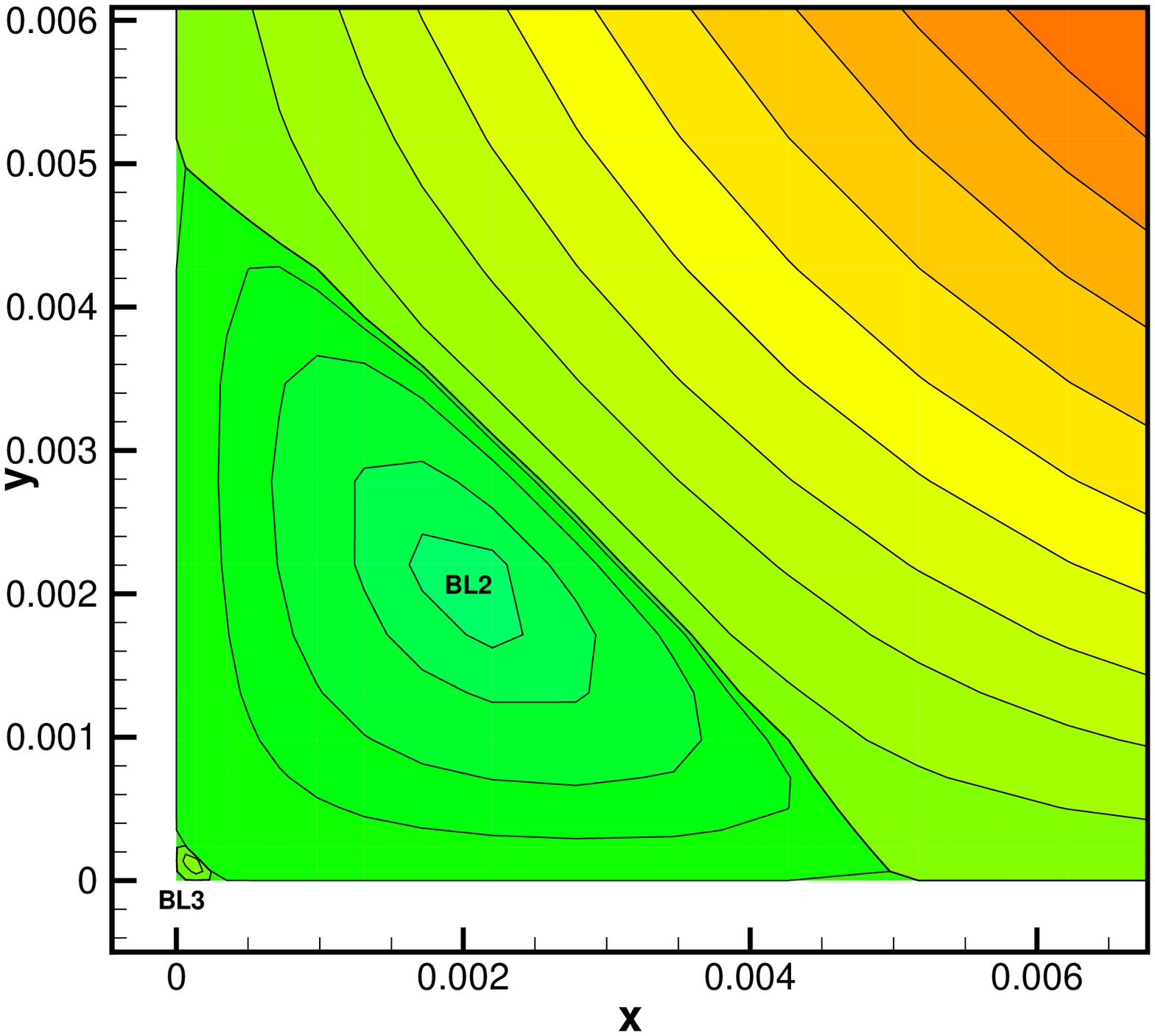}}
\end{minipage}
\begin{minipage}{.5\linewidth}
\centering{\includegraphics[width=5.65cm]{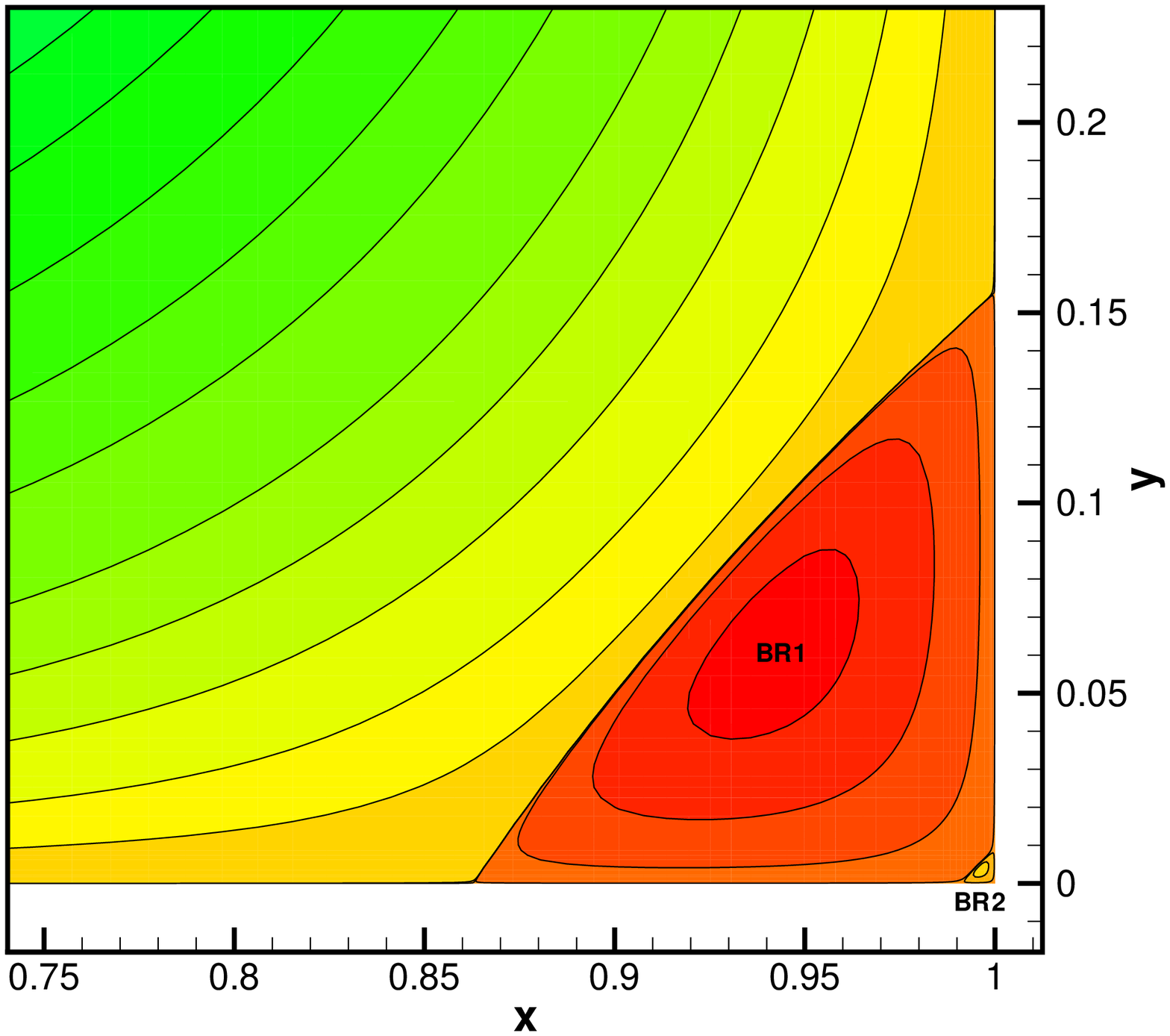}}
\end{minipage}
(b)
\begin{minipage}{.5\linewidth}
\centering{\includegraphics[width=5.65cm]{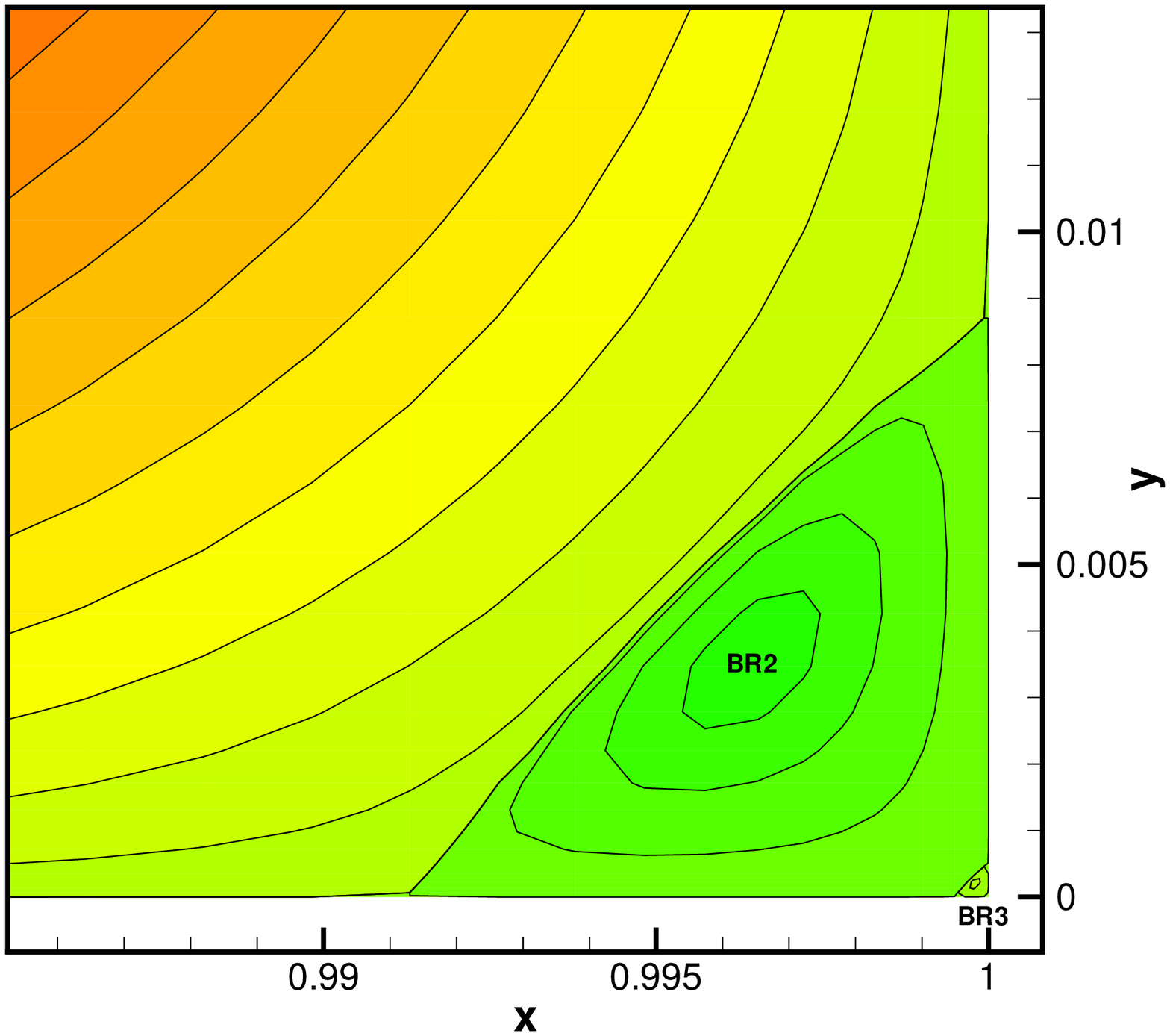}}
\end{minipage}
\caption{Moffatt vortices for $Re=100$ on grid size $161\times 161$: (a) BL1, BL2, BL3 at bottom left corner and (b) BR1, BR2, BR3 at bottom right corner.}\label{fig3}
\end{figure}

\begin{figure}[hH]
\begin{minipage}{.5\linewidth}
\centering{\includegraphics[width=5.65cm]{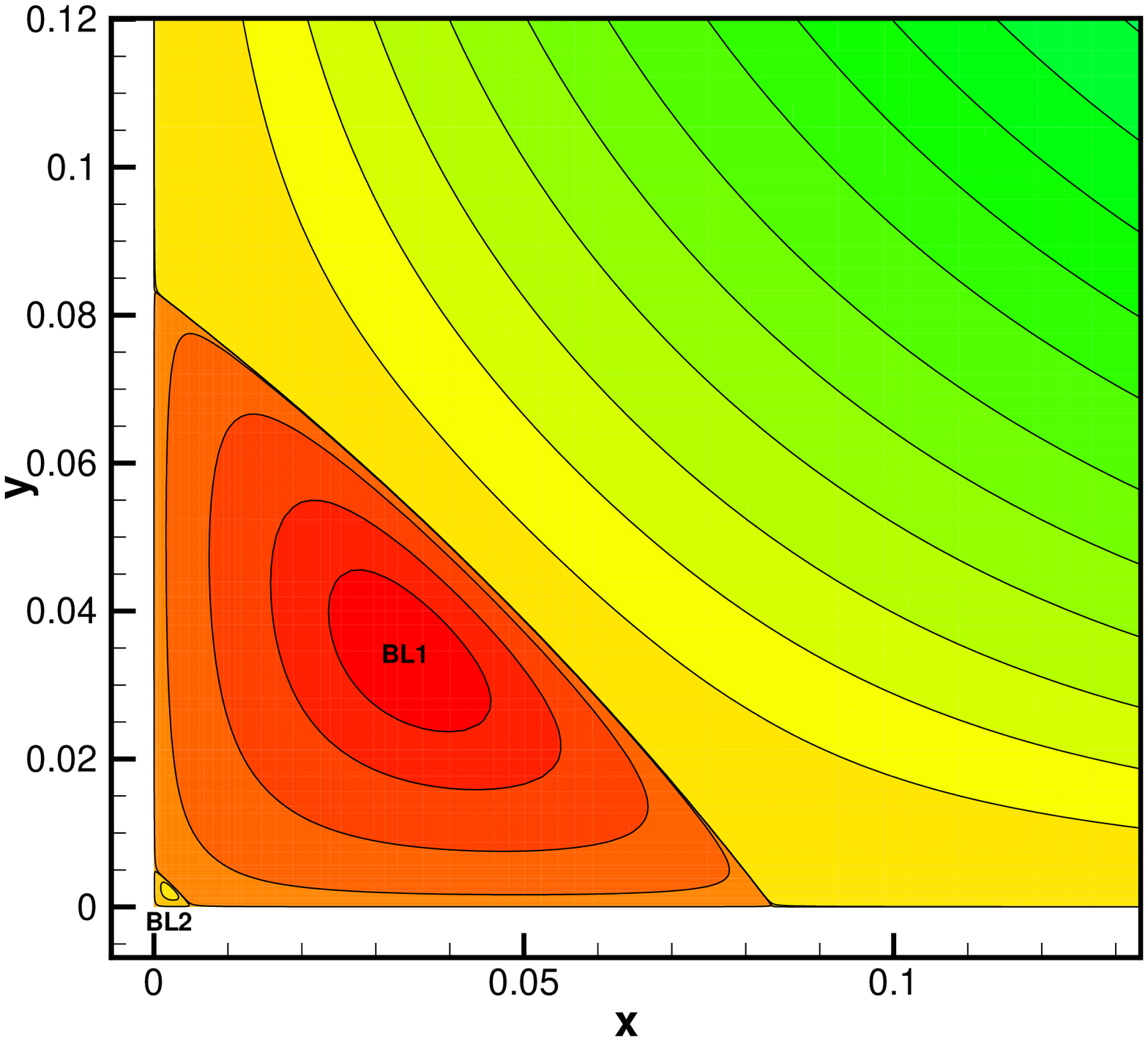}}
\end{minipage}
(a)
\begin{minipage}{.5\linewidth}
\centering{\includegraphics[width=5.65cm]{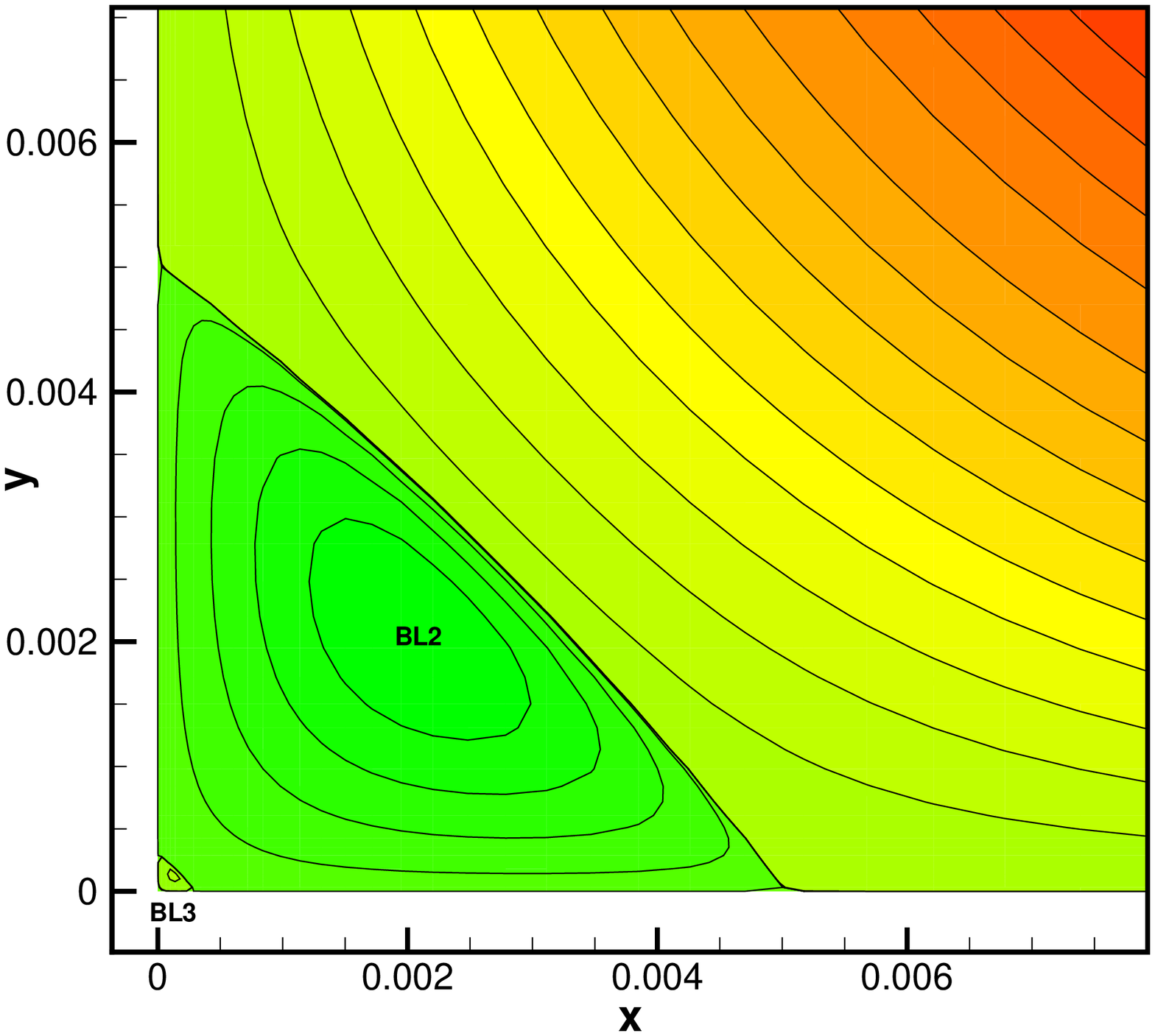}}
\end{minipage}
\begin{minipage}{.5\linewidth}
\centering{\includegraphics[width=5.65cm]{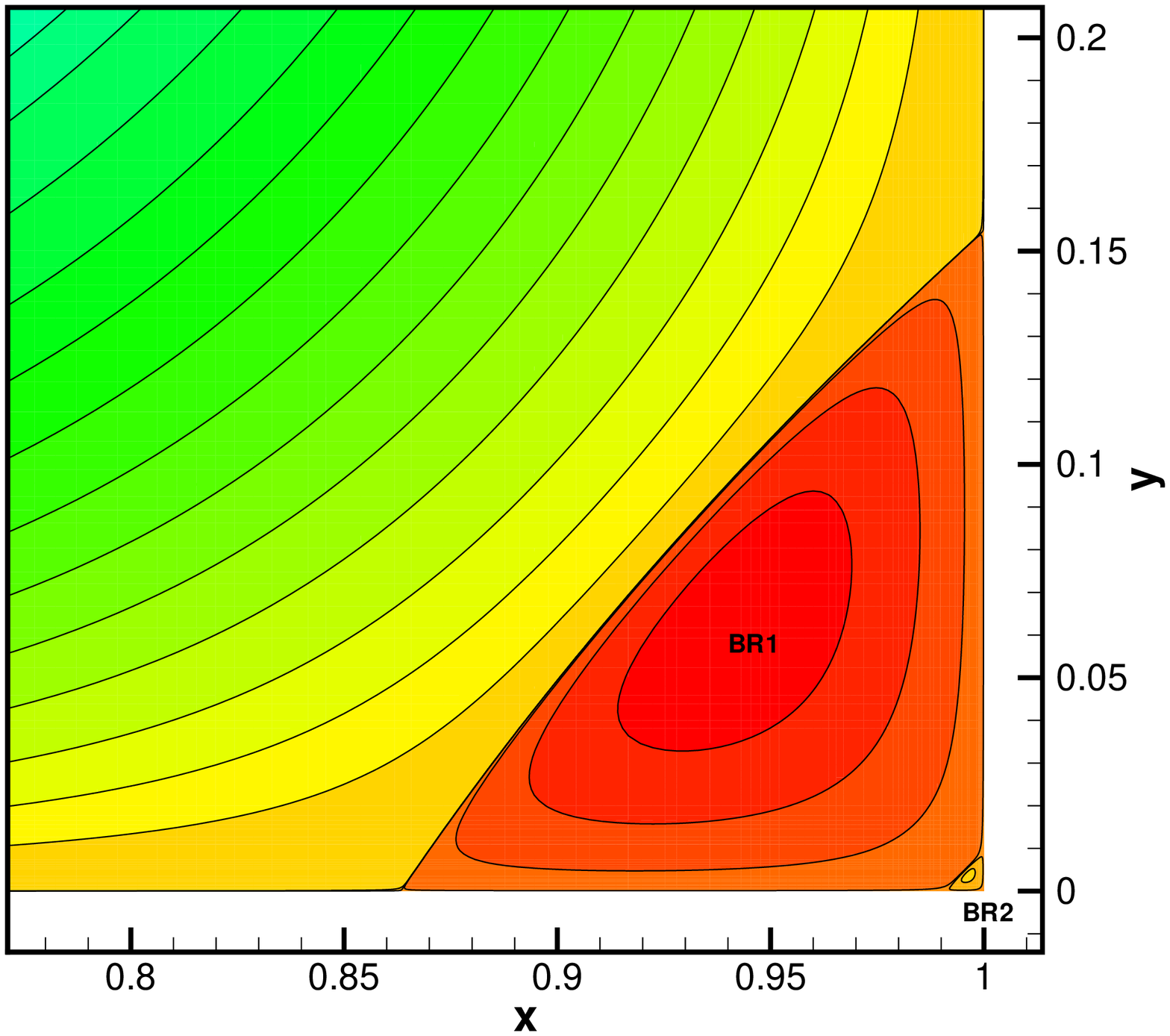}}
\end{minipage}
(b)
\begin{minipage}{.5\linewidth}
\centering{\includegraphics[width=5.65cm]{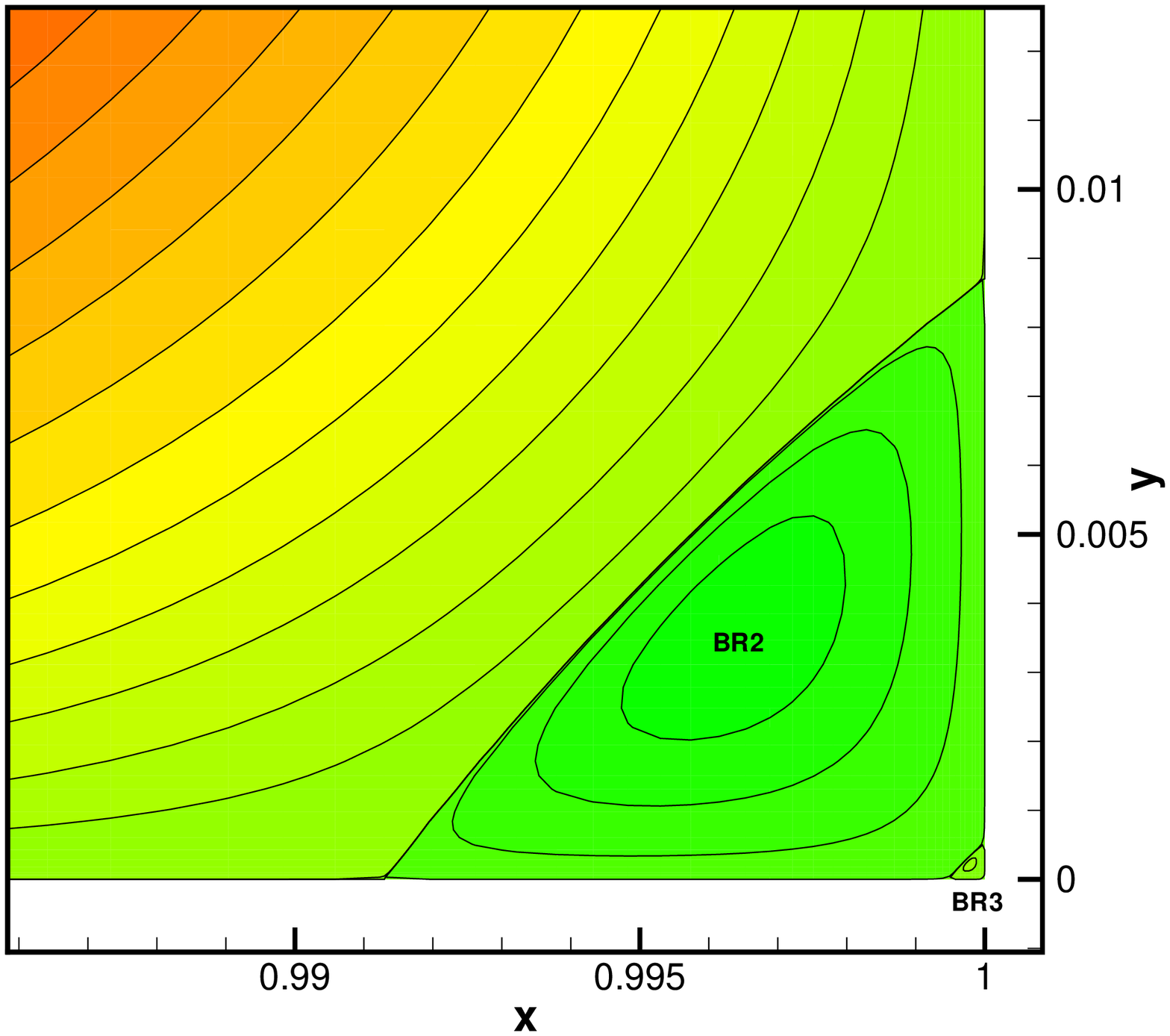}}
\end{minipage}
\caption{Moffatt vortices for $Re=100$ on grid size $321 \times 321$: (a) BL1, BL2, BL3 at bottom left corner and (b) BR1, BR2, BR3 at bottom right corner.}\label{fig4}
\end{figure}

It is a well known fact that even the best available computational resources are incapable of capturing all the members
of the so called infinite sequence of Moffatt vortices. With the computational resource available at our hand, we have been able to capture the first few members in that
sequence. We now present those members in terms of streamfunction contours (see Figure \ref{fig2}-\ref{fig4} ). For these vortices, the following nomenclature has been adopted. For the vortices at the
left corner, BL1 denotes the secondary vortex which is the first one to appear in the sequence. Likewise, BL2, BL3, BL4, $\cdots$ denote the tertiary, quaternary and
post-quaternary vortices respectively in that sequence in the same corner. In a similar way, for the Moffatt vortices in the right corner, the vortices in the same
sequence are denoted by BR1, BR2, BR3, BR4, $\cdots$ etc.
\clearpage
\subsection{Qualitative description of Moffatt vortices}
In Table \ref{tab1}, we provide properties of Moffatt vortices in terms of their intensities and sizes along with the location of their centers. The intensity of a vortex is defined by the streamfunction value at the center of the vortex. Following the work of H. K. Moffatt \cite{moff1, moff2}, we measure the size of a vortex in terms of the Euclidean distance between the center of the vortex and the corner of the cavity. We observe that the intensity as well as size of these vortices decrease rapidly, thus corroborating our findings with the theory. For a fixed Reynolds number, as grid size increases, one can observe the extreme closeness of the values of intensity and size in smaller scales also.
\begin{table}[hH]
\caption{Properties of Moffatt vortices for $Re=100$.}\label{tab1}
\begin{tiny}{
\begin{center}
\begin{tabular}{ccccc}
\hline
\\
& &\multicolumn{3}{c}{\bf Properties} \\  \cline{3-5}
\\
\textbf{Vortex} & \textbf{Grid size} &\textbf{Intensity} ($\psi$) & \textbf{Center location }($x$, $y$) & \textbf{Size}\\
\mr
BL1 &81$\times$81 &1.527017e-6 &(0.03469, 0.03469) &0.0490696\\
&161$\times$161 &1.799623e-6 &(0.03482, 0.03480) &0.0492360\\
&321$\times$321 &1.754731e-6 &(0.03470, 0.03471) &0.0490840\\
BL2 &81$\times$81 &-4.076319e-11 &(0.00220, 0.00221) &0.0031273\\
&161$\times$161 &-4.815543e-11 &(0.00222, 0.00221) &0.0031389\\
&321$\times$321 &-4.800864e-11 &(0.00220, 0.00220) &0.0031167\\
BL3 &81$\times$81 &--- &--- &---\\
&161$\times$161 & 1.262750e-15 &(0.00013, 0.00013) &0.0001971\\
&321$\times$321 & 1.212033e-15 &(0.00013, 0.00013) &0.0001958\\
\\ \hline
\\
BR1 &81$\times$81 &1.057025e-5 &(0.94104, 0.05955) &0.0838020\\
&161$\times$161 &1.260218e-5 &(0.94110, 0.06456) &0.0873908\\
&321$\times$321 &1.239697e-5 &(0.94321, 0.06179) &0.0839218\\
BR2 &81$\times$81 &-2.961589e-10 &(0.99655, 0.00349) &0.0049061\\
&161$\times$161 &-3.469230e-10 &(0.99654, 0.00348) &0.0049096\\
&321$\times$321 &-3.383937e-10 &(0.99653, 0.00347) &0.0049033\\
BR3 &81$\times$81 &--- &--- &---\\
&161$\times$161 & 9.422597e-15 &(0.99976, 0.00023) &0.0003274\\
&321$\times$321 & 9.047695e-15 &(0.99977, 0.00023) &0.0003259\\
\\ \hline
\end{tabular}
\end{center}
}
\end{tiny}
\end{table}
\\
In Table \ref{tab2}, we present the size and intensity ratio of two vortices in succession for step sizes $\frac{1}{81}, \frac{1}{161}, \frac{1}{321}, \frac{1}{641}$ and for zero-grid-step limit ($h \rightarrow 0$). The   ratios (both size and intensity) for step sizes $\frac{1}{641}$ and zero-grid-step limit are obtained by the Richardson's extrapolation and Lagrange's interpolation \cite{hoffman} of the computed data on coarser grids respectively.
\begin{table}[hH]
\caption{Intensity ratio and size ratio between two consecutive vortices for $Re=100$.}\label{tab2}
\begin{tiny}{
\begin{center}
\begin{tabular}{ccccccc}\hline
\\
& &\multicolumn{5}{c}{\bf Step Size ($h$)} \\  \cline{3-7}
\\
\textbf{Eddy Ratio} & \textbf{Properties} &1/81 &1/161 &1/321&1/641&$h\rightarrow0$ \\
\mr
BL1:BL2 & Intensity & 0.3746e5 & 0.3737e5 & 0.3655e5 & 0.3649e5 & 0.3667e5\\
&Size & 15.69063 & 15.68534 & 15.74826 & 15.75246 & 15.73847\\
BL2:BL3 & Intensity &---& 0.3813e5 & 0.3961e5 & 0.3970e5 & 0.3938e5\\
&Size &---& 15.92053 & 15.91181 & 15.91123 & 15.91317\\
\\ \hline
\\
BR1:BR2 & Intensity & 0.3569e5 & 0.3632e5 & 0.3663e5 & 0.3665e5 & 0.3658e5\\
&Size & 17.08097 & 17.79980 & 17.11507 & 17.06942 & 17.22158\\
BR2:BR3 & Intensity &---& 0.3681e5 & 0.3740e5 & 0.3743e5 & 0.3731e5\\
&Size &---& 14.99573 & 15.04196 & 15.04504 & 15.03477\\
\\ \hline
\end{tabular}
\end{center}
}\end{tiny}
\end{table}
\\
From Table 1 and Table 2, we observe that the intensity as well as the size for a particular grid-size go down in geometric progression with a common ratio which establishes H. K. Moffatt's theory \cite{moff1, moff2} accurately.
\subsection{Small scales resolution}
\begin{figure}[hH]
(a)
\begin{minipage}{.5\linewidth}
\centering{\includegraphics[width=6.5cm]{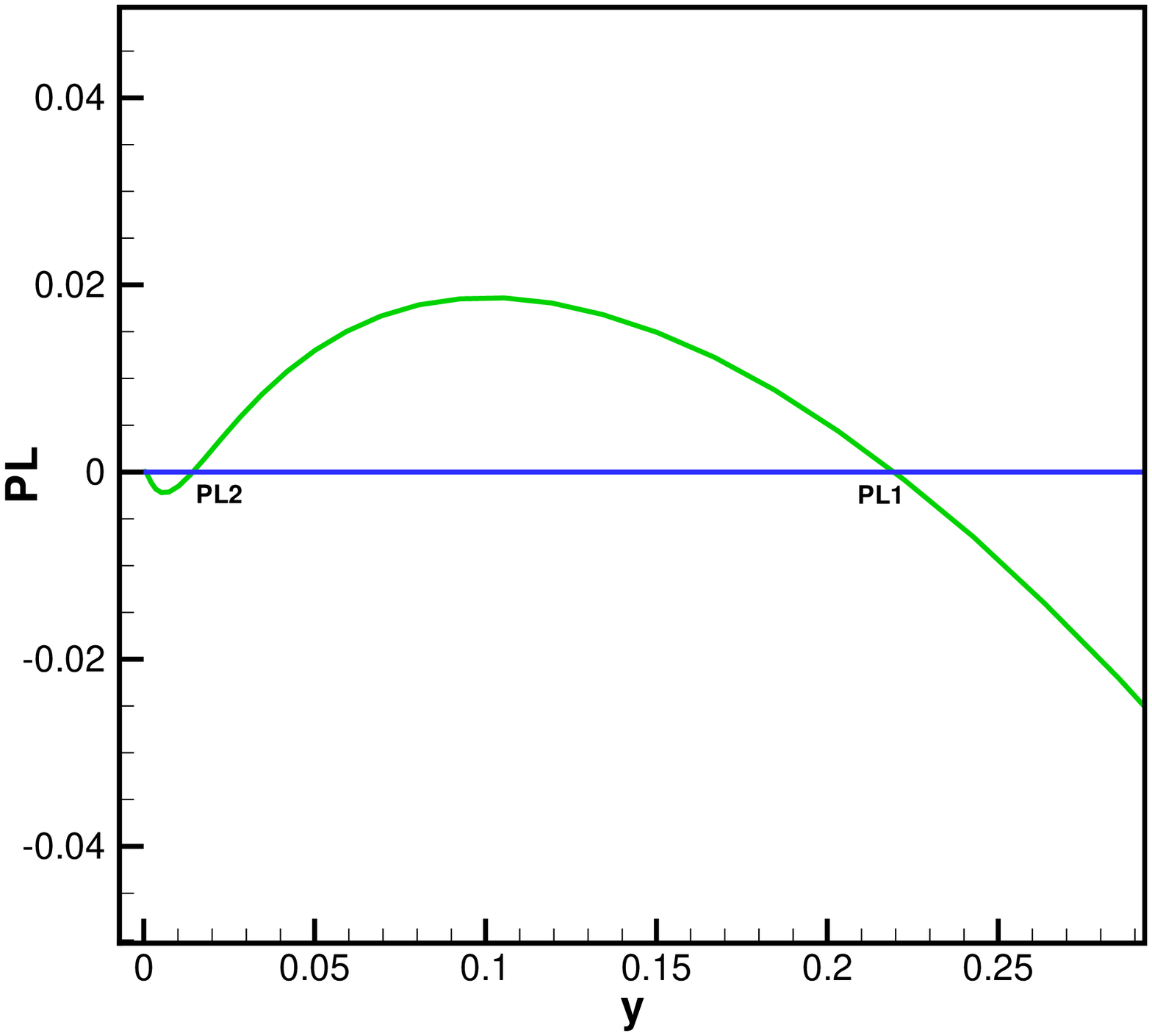}}
\end{minipage}
(b)
\begin{minipage}{.5\linewidth}
\centering{\includegraphics[width=6.5cm]{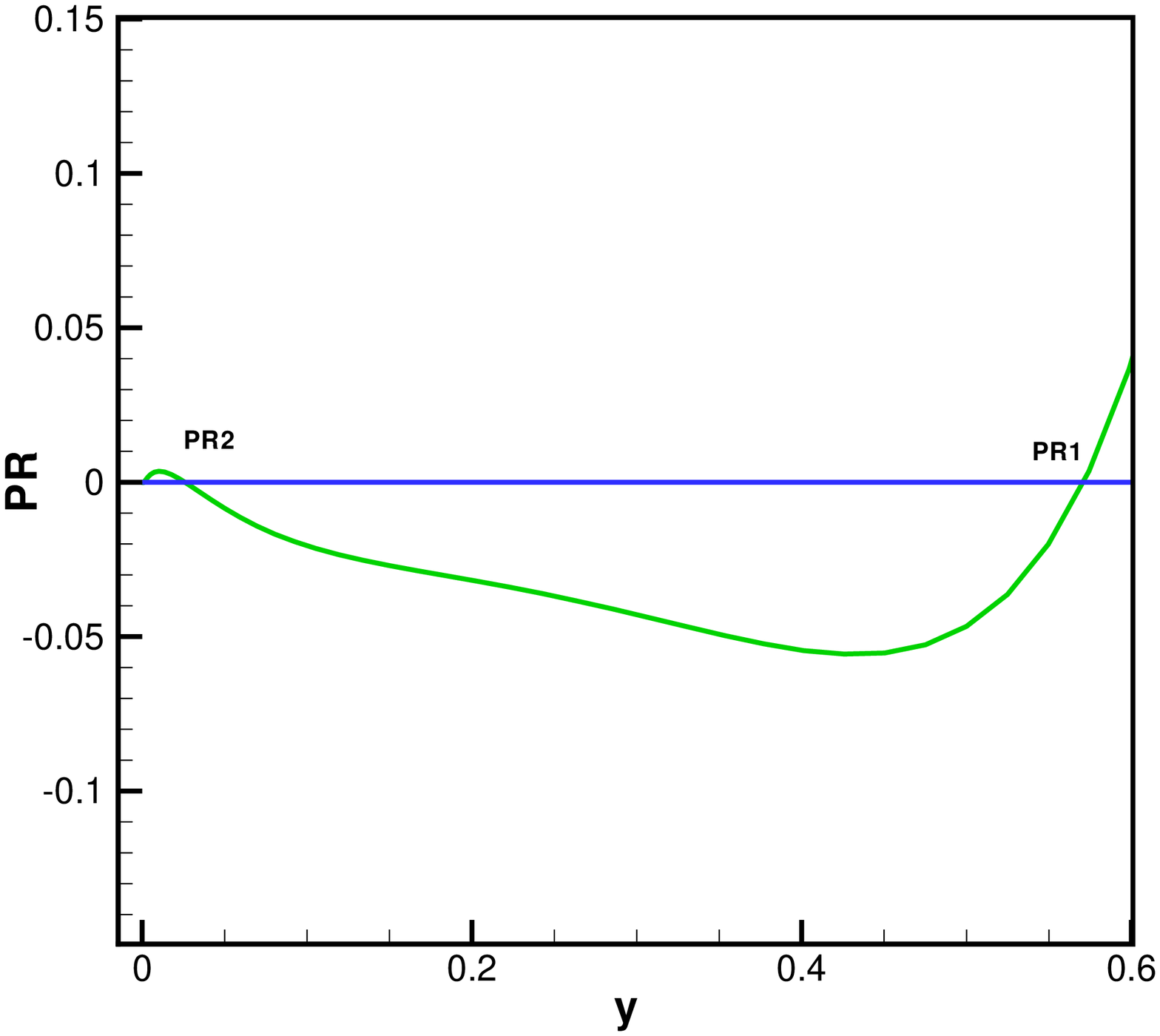}}
\end{minipage}
\caption{Pressure gradients along (a) left wall and (b) right wall  for $Re=100$ on grid size $81 \times 81$.}\label{fig5}
\end{figure}

By computing pressure gradients along the left and right walls of the cavity we conclude that our solution is not a spurious one. In Figure \ref{fig5}, we present changes in sign in pressure gradient in which PL1, PR2 denotes sign change from negative to positive and  PL2, PR1 denotes positive to negative.
It is well known that when pressure gradient changes sign along a wall, flow separation takes place and paves the way for the creation of a vortex. As such, the total number of vortices is equal to the total number of changes in sign in pressure gradient which establishes that our solution is accurate even in smaller scales and is free from numerical artifacts. 

 Similar facts can be observed for the grid sizes $161 \times 161$ and $321 \times 321$ and other Reynolds numbers considered here as well.
\section{Conclusion}\label{sec5}
We have established the existence of Moffatt vortices in the Lid-driven cavity flow by utilizing an HOC scheme at moderate Reynolds numbers, which, to the best of our knowledge has not been carried out before. The outcome of the computation of the ratio of size and intensity of these vortices strongly corroborates with the prediction of the theory of H. K. Moffatt. These has been further strengthened by the extrapolation and interpolation of the critical data.


\section*{References}
\begin{thebibliography}{12}
\bibitem{anderson} Anderson D M and Davis S H 1993 \emph{J. Fluid Mech.} {\bf 18} 1-31
\bibitem{col} Collins W M and Dennis S C R 1976 \emph{J. Fluid Mech.} {\bf 6} 417-432
\bibitem{hoffman} Hoffman J D 2001 \emph{Numerical Methods for Engineers and Scientists} (New York: Marcel Dekker)
\bibitem{jiten} Kalita J C, Dass A K and Nidhi N 2008 \emph{J. Comput. Appl. Math. } {\bf 214} 148-162
\bibitem{jiten2} Kalita J C 2007 \emph{Eng. Appl. Comput. Fluid Mech.} {\bf 1} 36-48
\bibitem{kelly} Kelley C T 1995 \emph{Iterative Methods for Linear and Nonlinear Equations} (Philadelphia: SIAM)
\bibitem{kirkinis} Kirkinis E and Davis S H 2014 \emph{J. Fluid Mech.} {\bf 746} R3
\bibitem{chetan} Malhotra C P, Weidman P D and Davis A M J 2005 \emph{J. Fluid Mech.} {\bf 522} 117-139
\bibitem{malyuga} Malyuga V S 2005 \emph{J. Fluid Mech.} {\bf 522} 101-116
\bibitem{moff1} Moffatt H K 1964 \emph{J. Fluid Mech.} {\bf 18} 1-18
\bibitem{moff2} Moffatt H K 1964 \emph{ Arch. Mech. Stosowanej} {\bf 2} 365-372
\bibitem{s1} Shankar P N 2005 \emph{J. Fluid Mech.} {\bf 539} 113-135
\bibitem{s2} Shankar P N 1993 \emph{J. Fluid Mech.} {\bf 250} 371-383
\bibitem{taneda} Taneda Sadatoshi 1979 \emph{J. Phys. Soc. Japan} {\bf 46} 1935-42
\end{thebibliography}
\end{document}